\documentclass[11pt]{article}
\usepackage{bm}

\usepackage[comma,super,compress]{natbib}

 \usepackage{graphicx}

\makeatletter
\renewcommand\@biblabel[1]{#1.}
\makeatother

\title{Local noncentrosymmetricity and possible spin-momentum locking in Sr$_3$Ru$_2$O$_7$ }

\author
{Chenyi Shen,$^{1,2,\dag}$ Hui Xing,$^{3,\dag}$ Xinxin Cai,$^2$ David Fobes,$^4$ Mingliang Tian,$^{5,6}$\\
 \\
Zhi-Qiang Mao,$^{4,\ast}$ Zhuan Xu,$^{1,6,\ast}$ and Ying Liu$^{2,3,6,\ast}$\\
\\
\normalsize{$^{1}$Department of Physics, Zhejiang University, Hangzhou 310027, China}\\
\normalsize{$^{2}$Department of Physics and Materials Research Institute,}\\
\normalsize{Pennsylvania State University, University Park, PA 16802, U.S.A.}\\
\normalsize{$^{3}$Key Laboratory of Artificial Structures and Quantum Control (Ministry of Education),}\\
\normalsize{Department of Physics and Astronomy, Shanghai Jiao Tong University,}\\
\normalsize{Shanghai 200240, China.}\\
\normalsize{$^{4}$Department of Physics, Tulane University, New Orleans, LA 70118, U.S.A.}\\
\normalsize{$^{5}$High Magnetic Field Laboratory, Chinese Academy of Science, Hefei 230031, China}\\
\normalsize{$^{6}$Collaborative Innovation Center of Advanced Microstructures, Nanjing 210093, China}\\
\normalsize{$\dag$These authors contributed equally}\\
\normalsize{$^\ast$To whom correspondence should be addressed; E-mail:}\\
\normalsize{mao@tulane.edu (Z-Q.M.) , zhuan@zju.edu.cn (Z.X.), and yxl15@psu.edu (Y.L.)}
}

\begin{document}
\date{}

\maketitle

\baselineskip24pt


\textbf{Strong spin-orbital coupling (SOC) was found previously to lead to dramatic effects in quantum materials, such as those
found in topological insulators. It was shown theoretically that local noncentrosymmetricity resulting from the rotation of
RuO$_6$ octahedral in Sr$_3$Ru$_2$O$_7$ will also give rise to an effective SOC\cite{SocSr327,MicroscopicnematicSr327}.
In the presence of a magnetic field applied along a specific in-plane direction, the Fermi surface was predicted to
undergo a reconstruction. Here we report results of our in-plane magnetoresistivity and magnetothermopower measurements
on single crystals of Sr$_3$Ru$_2$O$_7$ with an electrical or a thermal current applied along specific crystalline directions and
a magnetic field rotating in the $\bm{ab}$ plane (Fig. 1a), showing a minimal value for field directions predicted by the local
noncentrosymmetricity theory. Furthermore, the thermopower, and therefore, the electron entropy, were found to be suppressed
as the field was applied perpendicular to the thermal current, which suggests that the spin and the momentum in Sr$_3$Ru$_2$O$_7$
are locked over substantial parts of the Fermi surface, likely originating from local noncentrosymmetricity as well.}


Strontium ruthenates in the Roddlesden-Popper (R-P) series of Sr$_{n+1}$Ru$_n$O$_{3n+1}$ with a layered perovskite
structure\cite{RP} have attracted much attention since the discovery of superconductivity in the single-layer ($n$ = 1)
member of the series, Sr$_2$RuO$_4$~\cite{Sr214sc} and the subsequent demonstration of spin-triplet pairing in this
material\cite{Sr214Review1}. Work on other members of the series was motivated originally by the idea that the study of
these materials would provide insight into the mechanism of spin-triplet superconductivity in Sr$_2$RuO$_4$.
Interestingly, these other members were found to feature novel phenomena themselves.  In particular, the
bilayer member, Sr$_3$Ru$_2$O$_7$, was found to be characterized by an anisotropic first-order metamagnetic
transition\cite{metamagnetism}. When the field is along the $c$ axis, the metamagnetic transition was found below
$\sim$1 K around 8 T. The end point of the first-order phase transition line was found around 100 mK\cite{Fermiliquid,QCPThermodynamics},
leading to the presence of a quantum critical end point (QCEP)~\cite{metamagnetism,doublemetamagnetic}.
Near the QCEP, when the field was tilted slightly away from the $c$ axis, the resistivity obtained with the measurement
current perpendicular to the direction of the tilt is dramatically different from that obtained with the current parallel
with the tilt in a narrow field range, a finding attributed to the emergence of a nematic phase\cite{nematic}, Nematic Phase I,
which was tied to the presence of the QCEP. For the in-plane field, the metamagnetic transition was found below $\sim$1.2 K
around 5 T, very different from those found for the $c$-axis field. A QCEP was found to be present only under a
hydrostatic pressure\cite{In-pressureField1,In-pressureField2}. A large difference between values of resistivity obtained
with the in-plane field applied parallel with and perpendicular to the measurement current was again found, but over a very
large field range, and unrelated to the presence of QCEP. Even though this anisotropy was again linked to the presence of a
nematic phase\cite{nematic,nematic2}, Nematic Phase II, this phase must be very different from the Nematic Phase I because
the four-fold symmetry is already broken by the presence of the field. As a result, an anisotropy is expected to begin with.

The effect of the field orientation on metamagnetism in Sr$_3$Ru$_2$O$_7$ was attributed theoretically to a staggered SOC induced by local
noncentrosymmetricity, the breaking of local inversion symmetry due to the rotation of the RuO$_6$ octahedra about the $c$ axis for about
6.8 degrees in Sr$_3$Ru$_2$O$_7$, which separates Ru sites into A and B sublattices\cite{SocSr327}. This doubles the unit cell in real space,
or halves the first Brillouin zone in the momentum or $k$-space. Furthermore, the local noncentrosymmetricity also enables the hopping
between the $p_x$ and the $p_y$ orbitals of O ions that would have been forbidden in an undistorted tetragonal crystal. This SOC facilitates the
coupling of $t_{2g}$ orbitals of neighboring Ru ions which results in important features in the Fermi surface (FS) of Sr$_3$Ru$_2$O$_7$.
Interestingly, this effect can be described by Rashba-like spin-orbit coupling (SOC), $\lambda L_zS_z$, where $\lambda$ is the SOC
strength, $L_z$ the $z$-axis orbital angular momentum, and $S_z$ the $z$ component of the total spin\cite{SocSr327,MicroscopicnematicSr327}.
This picture explains the orientation dependence of metamagentism in Sr$_3$Ru$_2$O$_7$. Under a magnetic field
applied along the $c$ axis, both $L_z$ and $S_z$ remain good quantum numbers. As a result, the effect of the SOC would be limited
to the modification of the parameters in the Hamiltonian but not form of the Hamiltonian. Under the application of an in-plane
field, however, staggered magnetic moments are induced on the A and B sublattices featuring an opposite canting angle with
respect to the field direction, resulting in a net moment along the field direction and zero moment perpendicular to the field\cite{SocSr327}.

In the presence of an in-plane field, the local noncentrosymmetricity in Sr$_3$Ru$_2$O$_7$ will also lead to the reconstruction of the FS through
Pomeranchuk-type of effect, leading to corresponding change in the properties of the material\cite{SocSr327,MicroscopicnematicSr327}.  As shown in Fig. 1b, the FS of Sr$_3$Ru$_2$O$_7$ calculated in the tight-binding approximation and measured by angle resolved photo emission spectroscopy (ARPES) measurements\cite{ARPES} features six bands: one electron- (the $\delta$) and two hole-like (the $\alpha_1$ and $\alpha_2$) bands centered at the $\Gamma$ point, other two electron-like (the $\beta$ and $\gamma_1$) bands around the M point, and finally, a very small hole-like ($\gamma_2$) band near the X point. Among these bands, the $\gamma_2$ pockets were found to be extremely close to a van Hove singularity\cite{ARPES}. Under a sufficiently strong in-plane field applied along the $\Gamma-X$ direction, the two $\gamma_2$ pockets of the FS in the direction of the field were
predicted to be gapped out while each of the other two $\gamma_2$ pockets in the direction perpendicular to the field splits into two.
Therefore, within this picture, the sample resistivity and thermopower, which are dependent on the scattering rate and the entropy of the
charge carriers, respectively, would be expected to reach a minimum as the in-plane field is rotated to the $\Gamma-X$ direction.

We explored in the present work the effect of local noncentrosymmetricity by performing magnetoresistivity and magnetothermopower
measurements above the metamagnetic transition temperature on single crystals of Sr$_3$Ru$_2$O$_7$ cut along two separate crystalline
directions with the field rotating in the $ab$ plane. Specifically, a crystal of Sr$_{3}$Ru$_{2}$O$_{7}$ was cut along the $a$ (or $b$) axis
([100] direction) or 45$^o$ away from them
([110] direction), resulting in a 90$^o$- (the same as 0$^o$) or 45$^o$-cut crystal with a rectangular shape and measured by a steady-state
technique (Fig. 1a) to ensure that the applied electrical or thermal current is along the desired direction in real and momentum ($k$) spaces.
The direction of $-\nabla T$ relative to directions in the first Brilliouin zone is shown schematically by the arrows in Fig. 1b for the 45$^o$- and
90$^o$-cut crystals. It is known that charge carriers on different parts of the FS contribute to resistivity and thermopower differently, reflected
by a cosine of the angle between the $\vec{k}$ vector and the electric or thermal current, $\theta_{\vec{k}}$ (Supplementary Information 1).  For a 45$^o$-cut crystal, the electric or  thermal current is along $\Gamma-M$ direction, parts of the $\delta$, $\alpha_{1,2}$, $\gamma_1$ and $\beta$
sheets of the FS are perpendicular to (and the $\vec{k}$ vector parallel with) the electric or thermal current, making the cosine factor maximal ($\sim$1); for a 90$^o$-cut crystal, however, the electric or thermal current is along $\Gamma-X$ direction with parts of the $\delta$,
$\alpha_{1,2}$, and $\gamma_2$ sheets being perpendicular to the electric or thermal current. Therefore the 45$^o$ and 90$^o$-cut
crystals enable measurements on anisotropic transport properties arising from anisotropic FS: those FS parts perpendicular to the applied
electrical current or temperature gradient contribute the most to the sample resistivity or thermopower.

In Figs. 1c and d, thermopower, $S$, measured in $c$-axis oriented magnetic fields ($\mu_0H_\perp$) of zero and 9 T, is plotted against
the temperature, $T$. From 50 to 300 K, $S (T)$ of the 45$^o$-cut crystal was found to vary linearly with temperature, dropping notably
as the temperature fell below 50 K (Fig. 1c). In the 90$^o$-cut crystal, the linear behavior is not seen. Instead, $S (T)$ was found to vary slowly
for $T$ $>$ 50 K, dropping quickly again as the temperature fell below 50 K (Fig. 1d). In both crystals, $S (T)$ was found to show an increasing
rate of decreasing as the temperature was lowered passing 17 K. The linear behavior in $S (T)$ is observed in most metals and attributed to the
diffusion of free carriers\cite{Tpinmetals}. The behavior seen in the 90$^o$-cut crystal is clearly unconventional. Moreover, $S (T)$ is seen to
be suppressed by the application of $c$-axis field of 9 T below 50 K, which signals the magnetic fluctuation contribution to the thermopower.
In Figs. 1e and f, the Nernst signal, $e_y$, taken on crystals of both cuts with a 6 T magnetic field applied along the $c$ axis, is plotted as a
function of temperature. Above 50 K, the Nernst signal was found to be very small. The magnitude of Nernst coefficient $e_y$ is seen to
increase, reaching a maximum around 10 K, a temperature that is only slightly lower than the temperature at which a change of slope was
seen in $S(T)$. It is interesting to note that at 17 K, a peak was seen in the temperature dependent magnetic susceptibility\cite{Groundstate}
and Hall coefficients~\cite{hall} for Sr$_{3}$Ru$_{2}$O$_{7}$, attributed to the suppression of ferromagnetic fluctuation revealed by the inelastic
neutron scattering (INS) measurements\cite{NeutronZeroField,NeutronFiniteFields}.

The systematic behavior seen in the thermopower and Nernst effect measurements is striking. According to the
Heikes formula\cite{Hikes1,Hikes2},
\begin{equation}
S=\frac{\mu}{eT}=-\frac{\sigma}{e}
\end{equation}
where $\mu$ is the chemical potential and $\sigma$ is the entropy per electron, respectively. Therefore the thermopower measures
the entropy of charge carriers on the FS. Furthermore, according to the Mott formula~\cite{Mottform1,Mottform2}, the sign of $S$ for a
particular band is determined mainly by the slope of $A(E)$ where $A$ is the FS area. The ARPES studies showed that the FS sheets from the
$\alpha_{1,2}$ and $\gamma_2$ bands are hole-like, leading to a positive contribution to $S$ whereas Fermi surface sheets from the
$\delta$, $\beta$ and $\gamma_1$ bands are electron-like, resulting in a negative contribution to $S$. Given that a positive rather
than negative total $S$ was obtained for Sr$_3$Ru$_2$O$_7$ at all temperatures, contribution from holes on the $\alpha_{1,2}$ and
$\gamma_2$ sheets of FS should dominate the thermopower.

Measurements on the dependence of magnetothermopower on the angle between the in-plane magnetic field $H_{\parallel}$
and $-\nabla T$, $S (\theta)$, where $\theta$ = 0$^o$ corresponds to the case that the in-plane magnetic field is parallel with $-\nabla T$,
were carried out. $S (\theta$) taken on crystals of both cuts with an in-plane field of $\mu_0H_\parallel$= 6 and 9 T showed that
$S (\theta)$ possesses a four-fold symmetry, which is required by the crystalline symmetry. A significant observation, however, is that
the minimal values in the thermopower were found when the in-plane field was aligned along the $\Gamma-X$ direction,
as shown in Figs. 2a-d. Within the local noncentrosymmetricity SOC picture, two $\gamma_2$ pockets along the field direction were
gapped out so they no longer contribute to the thermopower, which will reduce the thermopower significantly despite of the small
size of the $\gamma_2$ pockets because of the large density of states resulting from the proximity to a van Hove singularity. The
other two $\gamma_2$ sheets that split into two under a high in-plane field will not contribute to thermopower at this field orientation
for the 90$^o$-cut crystal because the $\vec{k}$ vectors of these pockets are essentially perpendicular to the direction of the thermal
current at $\theta$ = 0$^o$. For the 45$^o$-cut crystal, the entropy from the surviving $\gamma_2$ pockets should be small because
of the band splitting. The SOC originating from the local noncentrosymmetricity should also have observable effect on the magnetoresistivity.
Indeed, the normalized resistivity, $\rho/\overline{\rho}_{max}$ shown in Figs. 3a-d revealed a four-fold symmetry, again as demanded by
the crystalline symmetry, and clear minimal value as the in-plane magnetic field is aligned along the $\Gamma-X$ direction, supporting the
prediction that two $\gamma_2$ pockets are indeed gapped out in this case, as expected in the local noncentrosymmetricity picture.

In addition to a four-fold symmetry and minimal values in the $\Gamma-X$ direction, a strong two-fold anisotropy was also observed in
$S(\theta)$. As seen in Fig. 2, at $T$ = 17 K, the thermopower is suppressed when the in-plane field is perpendicular to $-\nabla T$ for the 45$^o$-cut
crystal for $\mu_0H_\parallel$= 6 and 9 T, while for the 90$^o$-cut crystal, the opposite was found under the same condition. Similarly, two-fold
anisotropy was observed at $T$ = 20 K and several higher temperatures (Figs. 4a and b). Because of the presence of an in-plane field, the two-fold
anisotropy is also expected. Note that
the in-plane fields used here, $\mu_0H_\parallel$= 6 and 9 T, would have placed the system in the in-plane metamagnetic phase as
the critical field of the transition is around 5 T. However, the temperatures at which the measurements were carried out were all higher than
the critical temperature of the in-plane metamagnetic phase, $\sim$1.2 K. At the lowest measurement temperature, 2 K, even though the
two-fold anisotropy in $S(\theta)$ remains unchanged at 6 T, a dramatic change was found as the in-plane field was raised to 9 T.
For crystals of both cuts, a two-fold anisotropy featuring a strong suppression of thermopower by an in-plane field applied perpendicular
to $-\nabla T$ was found. The anisotropy change for the 90$^o$-cut crystal featuring a 180-degree flip is particularly striking. Similarly, a
two-fold anisotropy component is also seen in the normalized magnetoresistance, $\rho/\overline{\rho}_{max}$, but less anisotropic
than that seen in the thermopower, and only in the 90$^o$-cut crystal, as shown in Figs. 3a-d.

The above observation can be understood in the following scenario. Given that the two $\gamma_2$ pockets in the direction
of the field are gapped out, and  the other two $\gamma_2$ pockets do not contribute to thermopower or magnetoresistivity,
as described above, a two-fold anisotropy in $S(\theta)$
on crystals of both cuts, and the same in $\rho (\theta)$ on the 90$^o$-cut crystal, suggest that the entropy of charge carriers,
possibly the scattering rate as well, is reduced when a sufficiently strong in-plane field is applied perpendicularly to the momenta of the
charge carries on the FS, which in turn suggests that spins are locked perpendicularly to their momenta on substantial parts of the $\alpha_{1,2}$
sheets of FS. The application of an in-plane field along the direction of the locked spins would then tend to polarize the spin and reduce the entropy. Whether spins on the $\gamma_2$ pockets are
locked to the momenta is not clear as the contribution of the surviving $\gamma_2$ pockets to the thermopower is minimal because
of the cosine($\theta$)  factor mentioned above. The absence of the two-fold anisotropy
in $\rho (\theta)$ for the 45$^o$-cut crystal is understandable as the magnetoresistivity reflects the scattering of charge carriers as
well as the density of states, only the latter of which is tied directly to the entropy of a charge carrier.

The spin-momentum locking for carriers on the $\alpha_{1,2}$ sheets of the FS was not predicted in the local noncentrosymmetricity SOC
picture. In fact, the staggered magnetic moments induced by an in-plane magnetic field in the real space as predicted in this picture
would imply that spins should be aligned along the applied field direction over the entire FS in high in-plane magnetic fields. However,
in the zero or low in-plane field, spins on various sheets of the FS have their preferred direction because of the local noncentrosymmetricity
and orbital ordering as shown in the most recent theory of odd-parity electric octupole order featuring local electric quadrupoles arising
from the Pomeranchuk instability and/or the orbital order alternatively stacked in the bilayers\cite{OctupoleOrderSr327}. In addition, the
SOC from conventional sources, such as those found in Sr$_{2}$RuO$_{4}$\cite{Spin-obitalcoupling}, will also be present. It is entirely possible
that spins on certain parts of the $\alpha_{1,2}$ sheets of the FS would prefer to align perpendicularly to momenta.

An interesting question is whether the effect of local noncentrosymmetricity is temperature and (in-plane) field dependent. The line plots of
the angle dependent normalized thermopower obtained on both 45$^o$- and 90$^o$-cut crystals at various temperatures are presented
in Fig. 4a and 4b. Clearly the effect of the in-plane field on the thermopower as a function of temperature does not show a dramatic change
as the temperature was passing through 17 K. Similarly, no dramatic changes are seen as the system entered the metamagnetic state
(Figs. 4c and d). These observations suggest that spin-momentum locking in this material is insensitive to the temperature and the field,
consistent with the local noncentrosymmetricity picture.

The spin-momentum locking and the nematicity found in Sr$_3$Ru$_2$O$_7$ \cite{nematic,nematic2} may be related. Evidently the $c$-axis
nematic phase discovered originally is likely related to global physics such as Pomeranchuk type of Fermi surface instability near QCEP.
In contrast, the in-plane-field nematicity, which seems to be less unexpected as the four-fold symmetry is already broken by the presence
of the in-plane field, may be linked to local properties of the system such as the local noncentrosymmetricity and resulted SOC. Finally, it is
interesting to ask whether this effective SOC will lead to a topological phase. More theoretical studies are needed to understand our unexpected finding of the possible spin-momentum locking in Sr$_3$Ru$_2$O$_7$ .

\section*{Methods}
Single crystals of Sr$_3$Ru$_2$O$_7$ were grown by the floating zone method, as reported previously\cite{FZ}, and characterized by the x-ray
diffraction, magnetization, and electrical transport measurements to ensure their quality (Supplementary Information 2). No secondary phases
were detected in these screening measurements. Magnetoresistivity and thermoelectric measurements were carried out in a
Quantum Design PPMS-9 system (featuring a base temperature of 1.8 K) using a steady-state technique. In this setup, a thermal gradient, $-\nabla T$,
applied along the cut direction, 45$^o$ or 90$^o$ with respect to the $a$ or $b$ axis. The temperature gradient was determined by a pair of
differential type E thermocouples, and was set to around 0.5 K/mm.

\section*{Acknowledgements}
The authors have benefited from discussions with Hae-Young Kee and Yayu Wang. The work done at Penn State is supported by DOE under
DE-FG02-04ER46159, at SJTU by MOST of China (2012CB927403) and NSFC (11274229, 11474198 and 11204175), at ZJU by MOST of China
(2012CB927403) and NSFC (U1332209), at CAS by NSFC (U1432251), and at Tulane by NSF under DMR-1205469.

\section*{Author contributions}
C.S. and H.X. prepared the sample, performed the measurements, analyzed the data, and wrote the manuscript. X.C. perform the measurements and  analyzed the data. D.F. grew the crystals. M.T. performed measurements. Z-Q.M. and Z.X. analyzed the data and wrote the manuscript. Y.L. designed experiment, analyzed the data, and wrote the manuscript.

\section*{Competing financial interests}
The authors declare no competing financial interests.

\includegraphics[width=15cm]{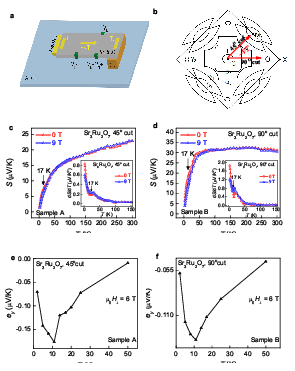}

\noindent \textsf{\textbf{Figure 1$\mid$ Experimental schematics and thermoelectric properties with magnetic filed along $\bm{c}$ axis.} \textbf{a,} The sample configuration for magnetothermoelectric measurements. \textbf{b,} Schematics showing the direction of $-\nabla T$ relative to
directions in the first Brilliouin zone for a 45$^o$- and 90$^o$-cut crystal, respectively. Here $\Gamma-X$ is along the direction
of Ru-O-Ru in real space; \textbf{c, d,} Thermopower, $S$ as a function of temperature ($T$) with a magnetic field ($H$) applied along the $c$ axis
for a 45$^o$- and 90$^o$-cut crystal, respectively. Inset: d$S$/d$T$ plotted against $T$, with a feature seen at 17 K; \textbf{e, f,} The Nernst signal, $e_y$ as a function of $T$ with a field, $\mu_0H_\perp$ = 6 T, applied along the $c$ axis for a 45$^o$- and 90$^o$-cut crystal, respectively.}

\includegraphics[width=15cm]{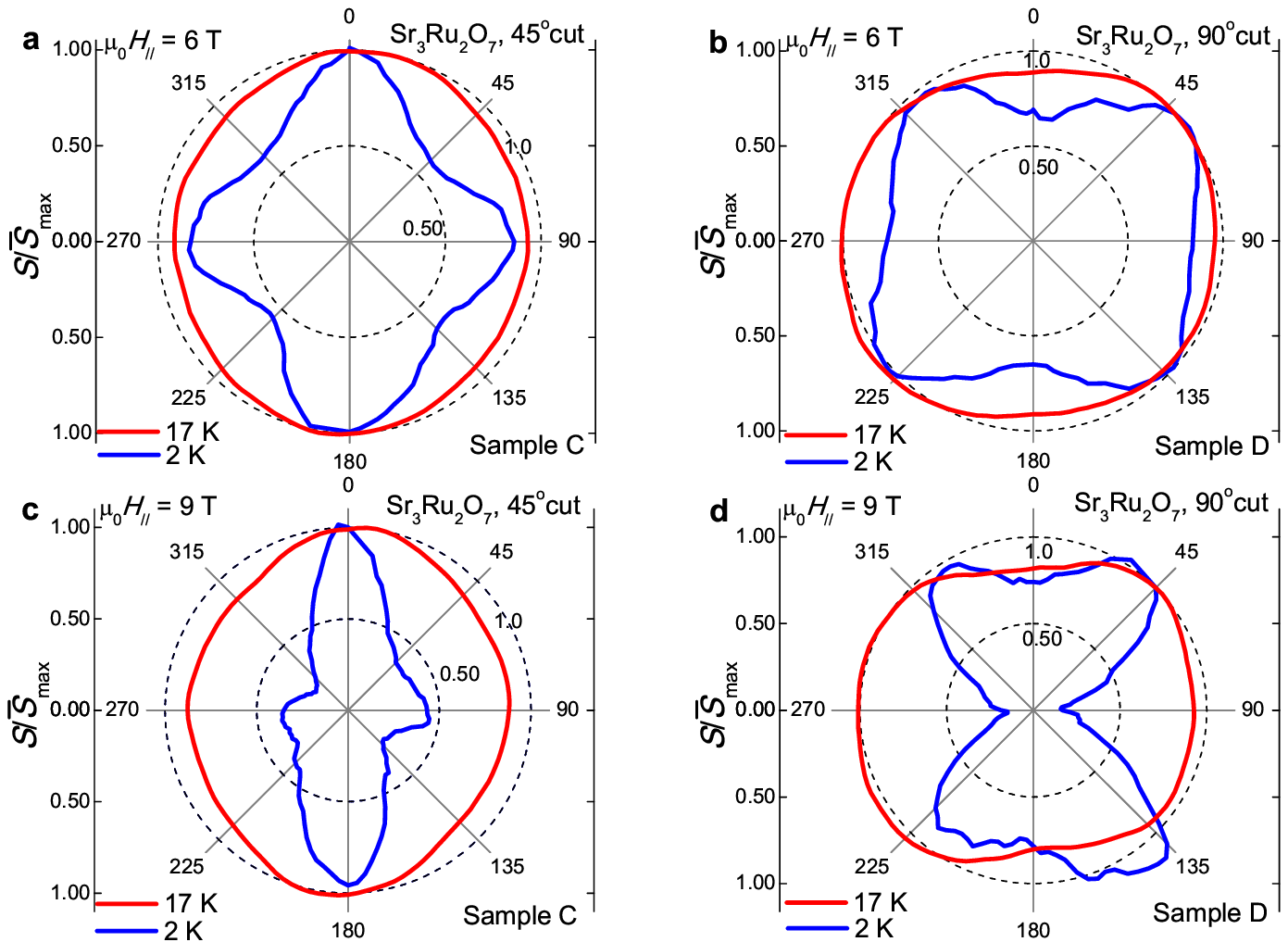}

\noindent \textsf{\textbf{Figure 2 $\bm\mid$ Normalized thermopower $\bm{S/\overline{S}_{max}}$ as a function of the rotating angle of the in-plane field.} \textbf{a, c} are for the 45$^o$-cut crystal; \textbf{b, d} are for the 90$^o$-cut crystal. Here $\overline{S}_{max}$, is the average of four maximal $S$ values and $\theta$ = 0$^o$ corresponds to the direction for which the magnetic field is parallel with $-\nabla T$, which is $\Gamma-M$ for the 45$^o$- and $\Gamma-X$ for the 90$^o$-cut crystal, respectively. The four-fold symmetry is expected from the crystalline symmetry. The two-fold anisotropy is also seen under certain conditions as noted. A minimum was observed in all cases when the in-plane field is placed along the $\Gamma-X$ direction.}

\includegraphics[width=15cm]{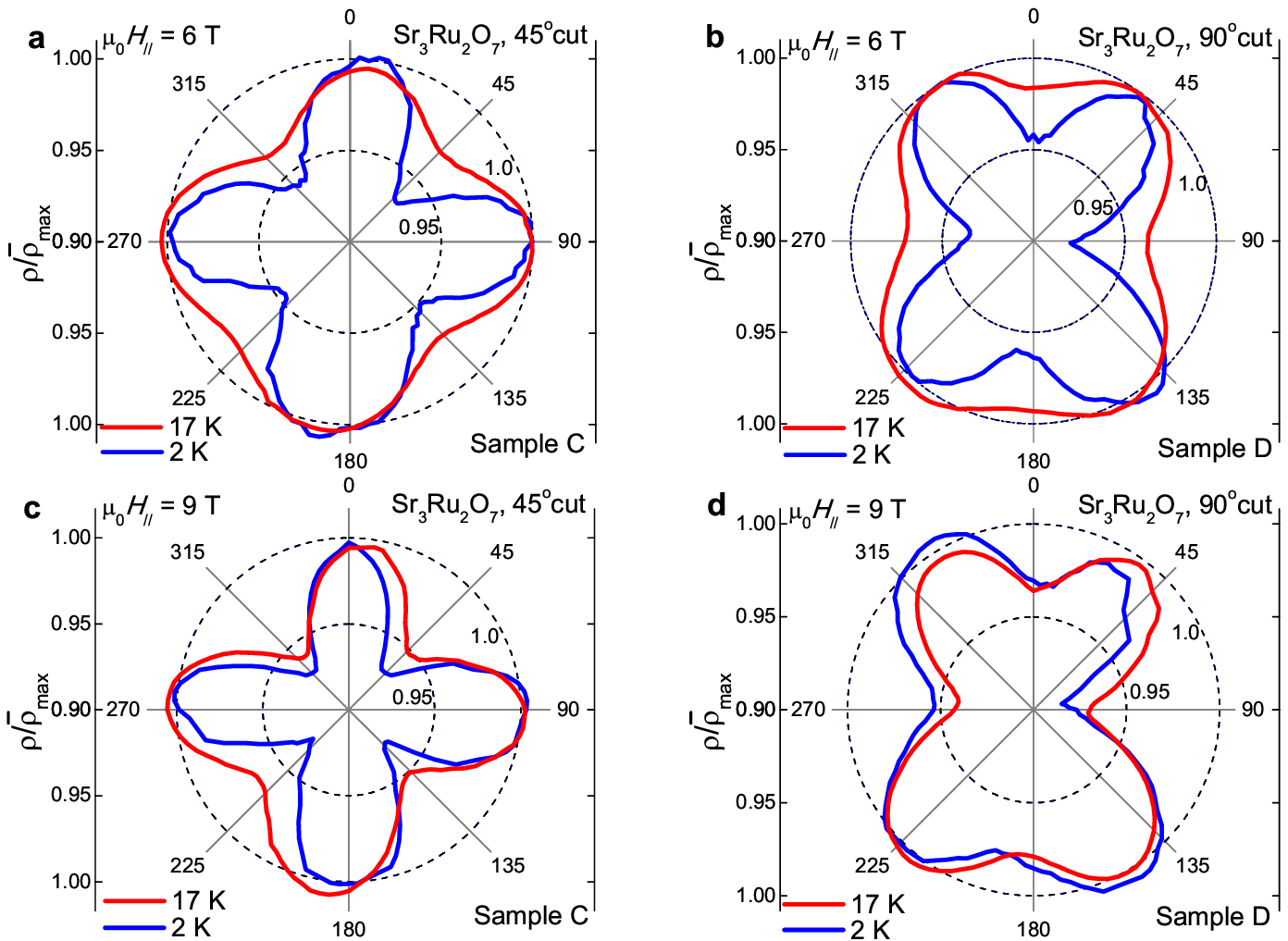}

\noindent \textsf{\textbf{Figure 3 $\bm\mid$ Normalized resistivity $\bm{\rho/\overline{\rho}_{max}}$ as a function of the rotating angle of the in-plane field.}  \textbf{a, c} are for the 45$^o$-cut crystal; \textbf{b, d} are for the 90$^o$-cut crystal. Two-fold anisotropy is also seen in $\rho/\overline{\rho}_{max}$ for the 90$^o$-cut (\textbf{b, d}) crystal, but less dramatic than that seen in $S(\theta)$.}

\includegraphics[width=15cm]{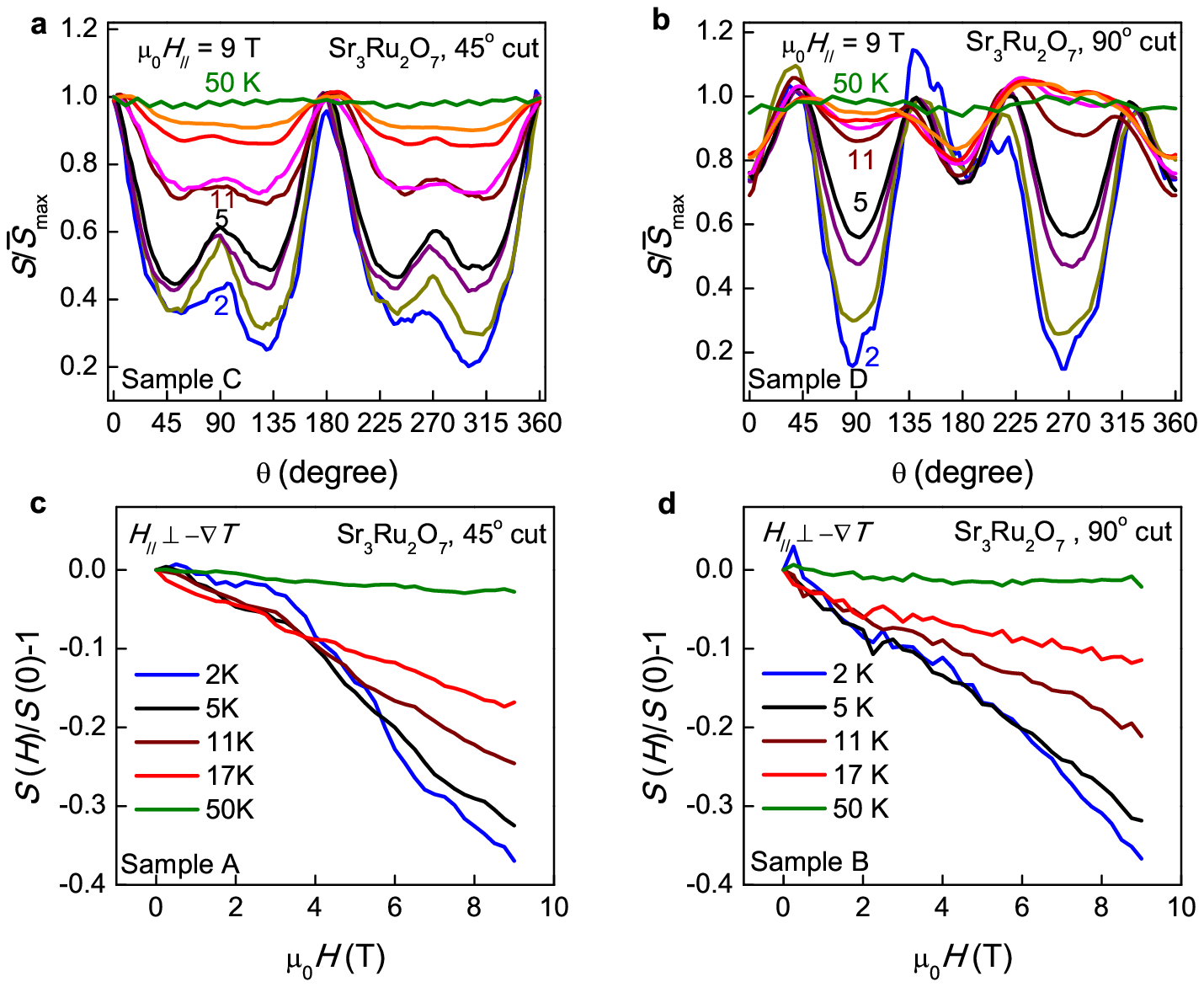}

\noindent \textsf{\textbf{Figure 4 $\bm{\mid}$ Temperature and field evolution of normalized thermopower.} \textbf{a, b,} Temperature evolution of normalized thermopower  $S/\overline{S}_{max}$ as a function of the rotating angle of a 9 T in-plane field for a 45$^o$- and 90$^o$-cut crystal, respectively. From bottom to top $T$ = 2 , 3, 4, 5, 11, 14, 17, 20 and 50 K. No abrupt change is seen in the anisotropy as the temperature is lowered; \textbf{c, d,} The normalized thermopower, $S(H)/S(0)-1$, as a function of the in-plane
magnetic field applied nearly perpendicularly to thermal gradient (within a few degrees) at fixed temperatures for a 45$^o$- and 90$^o$-cut single crystal, respectively.}

\end{document}